\documentstyle[12pt]{article}

\begin{document}

\title{Quantum Gravito-Optics: A Light Route from Semiclassical Gravity to
Quantum Gravity}
\author{Unnikrishnan. C S$^{1}$ and George T. Gillies$^{2}$ \\
$^{1}$Gravitation Group, Tata Institute of Fundamental Research, \\
Homi Bhabha Road, Mumbai - 400 005, India\\
$^{2}$School of Engineering and Applied Science, University of Virginia,\\
Charlottesville, VA 22904-4746, USA\\
$^{1}$unni@tifr.res.in, $^{2}$gtg@virginia.edu}
\date{Jl. ref: Class. Quantum Grav. 32, 145012 (2015), based on seed ideas
in our Gravity Research Foundation Essay 2013.}
\maketitle

\begin{abstract}
Quantum gravity remains an elusive theory, in spite of our thorough
understanding of the quantum theory and the general theory of relativity
separately, presumably due to the lack of any observational clues. We argue
that the theory of quantum gravity has a strong constraining anchor in the
sector of gravitational radiation ensuring reliable physical clues, albeit
in a limited observable form. In particular, all types of gravitational
waves expected to be observable in LIGO-like advanced detectors are fully
quantum mechanical states of radiation. Exact equivalence of the {\em full
quantum gravity theory with the familiar semiclassical theory }is ensured in
the radiation sector, in most real situations where the relevant quantum
operator functions are normal ordered, by the analogue of the optical
equivalence theorem in quantum optics.\ We show that this is indeed the case
for detection of the waves from a massive binary system, a single
gravitational atom, that emits coherent radiation. The idea of
quantum-gravitational optics can assist in guiding along the fuzzy roads to
quantum gravity.
\end{abstract}

Every physicist who has pondered over the fundamental theories of
gravitation and quantum mechanics is perhaps convinced that there should be
a theory of quantum gravity. Yet this theory that should incorporate the
fundamental principles of the two seminal theories has remained elusive in
spite of many serious efforts for more than half a century \cite{carlip}. \
Considering the fact that these principles -- the equivalence principle and
general coordinate invariance for the theory of gravitation and the
principle of superposition (uncertainty) and Lorentz invariance for quantum
(field) theory -- are simple to state and understand, this is very
surprising. \ The lack of any experimental clue whatsoever is usually
considered as the major impediment to finding a direction \cite{gill-unni}
and this indicates that the wait for a suitable theory might still be
painfully long.

However, when thinking along the lines of the development of quantum optics,
there are indeed some clues to quantum gravity, due to a fundamental
`optical' equivalence, familiar in quantum optics. \ The fundamental result
of quantum optics is that all states of light are fully quantum mechanical
and that they (their density matrix) can be represented as a diagonal
mixture of coherent states, with appropriate real weight factors acting the
role of distribution functions \cite{ecg1}. The coherent state that
represents `classical' monochromatic light itself is a superposition of all
number states \cite{glauber}, from $\left\vert n=0\right\rangle $ \ to $%
\left\vert n=\infty \right\rangle .$ Further, in those cases where the
relevant operators are normal ordered, the quantum and classical coherence
functions are exactly equal, without approximations \cite{ecg1}. The purpose
of this paper is to draw attention to a similar situation in gravity, by
explicitly showing that the coherent state of gravitational radiation
results from the emission of individual quanta even by a {\em single}
massive binary system. We will argue that this result is an {\em anchoring
constraint} on quantum gravity.

It is often argued that the unavoidable coupling between the quantized
matter and states of space-time is the strong reason for the inevitable need
for a theory of quantum gravity, since it would be inconsistent to consider
the states of matter as quantized while quantization is not reflected in the
associated field. \ A significant and relevant observation in this context
is that a sector of the gravitational field is in the form of radiation and
that all forms and states of gravitational radiation are inherently quantum
mechanical. \ The optical equivalence that we mentioned introduces a
fundamental equivalence between the quantum-signatured gravity and
semiclassical gravity in those cases where the operator order is normal --
not as an approximation, but {\em as an exact equivalence of observables}. \
The detection of gravitational waves \ belongs to this class because the
absorbed energy corresponds to the annihilation operators acting first.
Further, as we will show, emission by a massive binary system obeying
quantum dynamics is a quantum coherent state. In other words, any kind
gravitational radiation that is detected has to be a genuine quantum state
of radiative gravity. Therefore, when a full theory of quantum gravity is
formulated, its predictions for weak gravitational radiation, for which
nonlinear effects can be neglected, have to match {\em exactly }with the
results of semiclassical gravity in which only matter dynamics is initially
treated as quantum mechanical. To stress again the salient point, the
coherent radiation from the quantum mechanically treated dynamical system is
in fact quantum gravitational in an exact sense, physically and
mathematically. Further, the diagonal coherent state representation \ \cite%
{ecg1}, ensures that the radiation sector of the future quantum gravity
theory is already known formally and that it includes the coherent emission
from a single binary system.

The best representation of a `classical light wave' of constant
cycle-averaged intensity is the coherent state.\ However, the coherent state
is famously an elaborate quantum mechanical state, and indeed holds several
fundamental features of quantum optics \cite{glauber}. \ First of all it is
a superposition of {\em all number states} with Poissonian weights. \ It
embodies the uncertainty principle in a fundamental way, \ manifest through
its equal fluctuations in the phase-intensity quadratures observable in
`classical' experiments, without the need for photon counting, as intensity
fluctuations proportional to $\sqrt{I}$ with a white spectrum. Just as there
is really no classical light, there is no classical gravitational wave. Of
course, practical experiments today cannot detect the discreteness of quanta
inherent in such a wave due to the extremely large number density of
gravitons in even the weakest of gravitational waves, because of their low
specific energy, or energy per quantum $\hbar \omega $. \ Another important
difference compared to optics is the tiny coupling constant of the radiation
to the detectors. \ Therefore, in contrast to optical detectors with nearly
100\% efficiency, which can work with optical radiation of weak intensity
and hence large $\sqrt{I}$ fluctuations, gravitational wave detectors are
effectively `transparent'. \ Hence, detectable radiation fields contain a
huge number of gravitons and consequently the relative fluctuations, $1/%
\sqrt{I}$ $=1/\sqrt{N}$ approaches zero. The related, but different, issue
of the detectability of single gravitons has been discussed recently by F.
Dyson \cite{Dyson-IJMP13}.

The most important source for gravitational radiation detectable with
advanced detectors like LIGO and Virgo are orbiting binary pulsars. The
dynamics of such systems within usual quantum (Schrodinger) theory is that
of two particles in a central gravitational potential, even though we deal
with it usually as a Newtonian problem, under suitable approximations,
because the relevant action is much larger than the Planck constant. We note
that there is now direct experimental evidence, from experiments with cold
neutrons, that particle dynamics in a Newtonian gravitational field has
quantized energy levels and observables, as predicted by the Schrodinger
equation \cite{Neutrons} and a binary neutron star system is no exception,
despite its mass and size. First we show the main result of the paper, that
gravitational radiation emitted by a {\em single} massive binary system
within quantum theory is indeed the quantum coherent state of radiation
consisting of a superposition of all number states with Poissonian weights,
and then discuss the general implications. This might seem surprising at
first sight, since the usual coherent state in quantum optics is typically
generated by an ensemble of population-inverted atoms undergoing stimulated
emission, as in \ a laser, which locks the phase of the emitted wave. How is
it possible that the spontaneous emission from a single gravitational atom
-- two massive objects in orbits around each other -- be a coherent state?

Two \ massive objects in a quasi-stationary orbit indeed exhibit very
closely spaced Bohr-like orbital behavior at any instant, as described by
the basic quantum mechanical treatment of two particles bound by the long
range Coulomb-like gravitational force. The discrete, yet seemingly and
practically continuous energy states of such a system is approximately 
\begin{equation}
E_{n}=-\frac{\left( GM_{1}M_{2}\right) ^{2}\mu }{2\hbar ^{2}n^{2}}
\end{equation}%
quite in analogy with quantized orbits of charged particles. $\mu $ is the
reduced mass and approaches $M/2$ for nearly equal massive bodies. Since the
total energy is $E=-GM_{1}M_{2}/2r,$ we get 
\begin{equation}
n^{2}=GM_{1}M_{2}\mu r/\hbar ^{2}\simeq GM^{3}r/2\hbar ^{2}
\end{equation}%
The quantum number $n$ is of course extremely large for the massive binary
neutron star system, of the order of $10^{76}$ even at the last stages of
collapse. Nevertheless the energy levels are quantized in standard physical
theory that expects quantum theory to be applicable to all material bodies,
and in particular for this system in orbit under a long range central force.
Changes in the orbit can occur only by radiation of single quanta, due to
the coupling to the enormous number of open external degrees of freedom, or
the gravitational environment. Since $n$ is very large and both masses are
taken to be nearly equal for approximate estimates, 
\begin{equation}
\frac{\Delta E}{\Delta n}\simeq \frac{dE}{dn}=\frac{G^{2}M^{5}}{2\hbar
^{2}n^{3}}
\end{equation}%
Therefore, 
\begin{equation}
\Delta E\simeq \frac{GM^{2}}{rn}\Delta n=\hbar \left( \frac{2GM}{r^{3}}%
\right) ^{1/2}\Delta n=\hbar \omega
\end{equation}%
where $\omega $ is the frequency of the emitted quantum. This is of course
time dependent, but quasi-constant on the time scale of emission of a large
number of quanta because $\Delta E$ is typically smaller than $10^{-30}$ J
and the power in gravitational wave emission is larger than $10^{20}$ W even
in the early stages of binary evolution. Each quanta corresponds to $\Delta
n=2,$ to conform to the quadrupole character of emission and its selection
rule of change of angular momentum by $2\hbar $ in each transition ( the
states of the binary system are such that the angular momentum quantum
number $l\simeq n$). Since the orbital frequency of the binary is $\Omega
=\left( 2GM/r^{3}\right) ^{1/2}$ we get $\Delta E=\hbar \omega =2\hbar
\Omega $ and the frequency $\omega $ of the emitted quantum `graviton' is
twice the orbital frequency $\Omega .$ (In this case, this is perhaps a
coincidence specific to the particular assumptions made, but we expect this
relation in general. Also, the fact that the frequency associated with the
quanta emitted in a very short section of the orbit encodes the information
of the full orbit in terms of its period is not surprising because we are
dealing with smooth circular orbits for which a part encodes the whole). In
this scenario, we expect decay of the orbit by spontaneous emission of the
quanta. The instant of emission is completely random for this single system,
as in all such quantum mechanical systems, because there is no possibility
of stimulated emission.

For slowly varying amplitude of the radiation (cycle average intensity $\bar{%
I}$ slowly varying), we note that the average number of quanta $\bar{m}$ in
a short fixed duration $T$, which is small compared to the orbital time but
long compared to the lifetime of single energy eigenstates,\ is 
\begin{equation}
\bar{m}=\bar{I}T/\hbar \omega
\end{equation}%
$\bar{m}$ is a very large number even for very small time intervals since $%
\Delta E=\hbar \omega $ is miniscule and the power radiated in gravitational
waves is enormous in comparison -- $P(\omega )\sim 10^{25}-10^{50}$ W during
the observable gravitational evolution of such binary systems, for example.
Given that the process of emission is random spontaneous emission, it can
occur at any instant in the time interval $T.$ We assume that the
probability of multiple emissions at any instant is zero, which is
reasonable for the single binary system. So, the probability of emission of $%
m$ quanta is the Poissonian distribution, 
\begin{equation}
P(m,T)=\frac{\bar{m}^{m}}{m!}\exp (-\bar{m})
\end{equation}%
with a probability amplitude of 
\begin{equation}
A(m,T)=\frac{\bar{m}^{m/2}}{\left( m!\right) ^{1/2}}\exp (-\bar{m}/2)
\end{equation}%
In a given duration $T$ of the order of an orbital cycle, the quantum
process is determined by the sum of amplitudes to emit different number of
quanta, each in a quantum state $\left\vert m\right\rangle ,$ with the
average number of quanta emitted over an interval $T$ short compared to the
orbital time constrained to be $\bar{m}=\bar{I}T/\hbar \omega .$ The quantum
state of the radiation is the superposition of all these states $\left\vert
m\right\rangle $ with corresponding amplitudes $A(m,T)$. Noting that the
emitted quanta are bosonic, each term simply adds in phase in the
superposition. We thus get the final quantum state of gravitational
radiation from a single massive binary system or a gravitational atom as, 
\begin{equation}
\left\vert \alpha \right\rangle =\exp (-\bar{m}/2)\sum\limits_{m=0}^{\infty }%
\frac{\bar{m}^{m/2}}{\left( m!\right) ^{1/2}}\left\vert m\right\rangle =\exp
(-\left\vert \alpha \right\vert ^{2}/2)\sum\limits_{m=0}^{\infty }\frac{%
\alpha ^{m}}{\left( m!\right) ^{1/2}}\left\vert m\right\rangle  \label{state}
\end{equation}%
where we have written $\bar{m}=\left\vert \alpha \right\vert ^{2}.$ This is
the same as the coherent state in quantum optics. It is given by 
\begin{equation}
\left\vert \alpha (t)\right\rangle _{\omega }=\exp (a^{\dagger }\alpha
-\alpha ^{\ast }a)\left\vert 0\right\rangle _{\omega }=\exp (-\left\vert
\alpha \right\vert ^{2}/2)\sum\limits_{n=0}^{n=\infty }\frac{\alpha ^{n}}{%
\sqrt{n!}}\left\vert n\right\rangle _{\omega }  \label{coherent}
\end{equation}%
obeying the fundamental quantum equation $a\left\vert \alpha \right\rangle
=\alpha \left\vert \alpha \right\rangle ,$ where $a$ is the relevant
annihilation operator. We have indicated that the amplitude is time
dependent in general, and the actual waveform itself admits a further
superposition over different frequencies, $\omega .$ \ The expectation value
of the gravitational field in such a state is 
\begin{equation}
\left\langle \alpha \right\vert \hat{E}_{g}\left\vert \alpha \right\rangle
=\left\vert \alpha \right\vert ^{2}\sin \omega t=\left\vert \alpha
\right\vert ^{2}\sin 2\Omega t
\end{equation}%
indicating the harmonic `classical' character. However, the fluctuations in
amplitude and phase obey the uncertainty relation, and hence the state is
fully quantum mechanical.

We have shown that the detailed quantum state of gravitational radiation
emitted by a\ single massive binary system is a superposition of all number
states, including the null (vacuum state) since no gravitational quanta
might be emitted in any given time interval. {\em Thus the coherent
gravitational wave that we expect to detect soon with ground based detectors
is in fact the quantum gravitational coherent state}. This state has
expectation values that are sinusoidal in time in the two quadratures, with
a stable phase, because it is always an eigenstate of the annihilation
operator with its time dependence $\hat{a}(t)=\hat{a}(0)\exp (-i\omega t).$
This is consistent with the physical feature that emission of the low energy
quanta does not affect the orbital phase of the binary system during the
short interval $T,$ which is much smaller than the orbital period. A notable
feature in our demonstration is that multiple spontaneous emission events in
a single physical system -- the binary system -- over an extended temporal
interval results in the same coherent state radiation expected from the
stimulated emission from a statistical ensemble (multiple systems and
multiple events at an instant, as in many atoms emitting coherent photons in
a laser medium). This result relies on the fact that we are dealing with
very high quantum number Rydberg-like gravitational orbits which are
susceptible to even the smallest of external perturbations (of the order of
energy level separation) and there are a large number of quanta emitted in a
duration much smaller than the orbital period. Our result applies also to
atomic systems in very large-n orbits where the electromagnetic radiation is
expected to be close to the coherent state in principle, resembling
classical radiation from a charged particle moving in a large size orbit. We
note in passing that there is no chance of seeing the discreteness of
emission in the gravitational case, in contrast to the case of a single
atomic system, because one remains in the very high quantum number situation
even at the point of coalescence since the `Bohr radius' in this case, of
order $\hbar ^{2}/GM^{3}$, is much smaller than even the Planck length.

A brief discussion about the corrections to this scenario arising from
finite size effects, tidal distortions, relativistic effects, spin etc. is
in order. All these aspects affect the time evolution of the orbit and
therefore the amplitude and the frequency of the gravitational waves. In the
classical theory this is handled by perturbation theory by suitably adding
relevant post-Newtonian terms. This changes only the effective potential of
the problem, perturbatively, but neither the fundamental quadrupolar nature
of the weak emission nor the fact that the energy levels are quantized. The
fundamental relation $E=\hbar \omega $ for the emitted quanta remains valid
at each instant, with $\omega $ only perturbatively different from the
simple Newtonian situation. Hence, none of these corrections affect the
picture of emission of a large number of quanta at nearly constant frequency
and constant probability factors in equation (6) when we consider time
intervals that are orders of magnitude smaller than the orbital frequency $%
\Omega .$ Even in the smallest imaginable duration comparable to Planck
time, millions of quanta are emitted from an early binary system because the
average $\bar{m}$ is given by $P(\omega )/\hbar \omega .$ Hence, all time
dependence can be neglected for the derivation of the coherent state
emission from the binary system and the equations (6-10) remain fully valid.
Slow time dependences will then be reflected in the slow variation of $%
\omega $, the amplitude operator of the field $\hat{E}_{g}$ and in the
overall spectrum of the emitted gravitational waves. In other words, the
familiar post-Newtonian and tidal corrections will be reflected in the
expectation value of the waveform in the quantum treatment, in accordance
with the Ehrenfest theorem.

Many phenomena in quantum optics have their expectation values exactly
matching results of semi-classical optics, in which light is considered as
classical waves and matter is fully quantized. \ The celebrated optical
equivalence theorem \cite{ecg1} guarantees this equivalence when the
relevant operator functions are normal ordered, with all the creation
operators arranged to the left of the annihilation operators. \ In other
words, what we call classical light is in fact quantum light with certain
specific relationships to the matter that generates it. \ The
quantum-gravitational optical equivalence involving gravitational waves and
their matter sources and detectors implies that the results of the full
theory of quantum gravity in its radiation sector with observables
represented by normal ordered operators will in fact be identical to the
results in semiclassical gravity. \ This insight is a substantial
improvement over the current situation with no theory of quantum gravity, to
a phase where we can rely on a limited theory of quantum gravity for
phenomena involving gravitational waves and matter in arbitrary quantum
states. What is perhaps excluded in such a framework are those quantum
states of space-time where the characteristic nonlinear nature of the
gravitational interaction would play a role.

We now summarize the main results of {\em quantum-gravitational optics}.

1) Gravitational waves from a single massive binary system obeying ordinary
quantum mechanics is in a quantum coherent state.

2) Representation of arbitrary graviton states in a diagonal coherent state
representation \cite{ecg1}: It is indeed remarkable that we already know how
to represent the arbitrary quantum states of quantized gravitational
radiation in a complete theory of quantum gravity -- since the coherent
states $\left\vert \alpha \right\rangle $ are quantum mechanical \cite%
{glauber} and overcomplete, {\em any} quantum state (density matrix) of
gravitational radiation can be represented in a diagonal coherent state
representation \cite{ecg1}, 
\begin{equation}
\hat{\rho}=\int \phi (\alpha )\left\vert \alpha \right\rangle \left\langle
\alpha \right\vert d^{2}\alpha
\end{equation}%
with $\left\vert \alpha \right\vert ^{2}=\bar{n}=\bar{I}/\hbar \omega $ and $%
Tr(\hat{\rho})=\int \phi (\alpha )d^{2}\alpha =1,$ where $\phi (\alpha )$
are real distribution functions (and not operators). This brings out the
important feature of quantum gravitational optics and quantum gravity, as in
familiar quantum optics, that all radiation states are quantum states. \ 

3) The gravito-optical equivalence theorem, detection and correlations: The
emission and detection of gravitational waves are described in quantum
gravity with non-commuting creation and annihilation operators, symbolically
written here as $a^{\dagger }$ and $a$. \ The diagonal coherent state
representation \cite{ecg1} implies that for all instances of gravitational
radiation and detection, the quantum gravity prediction for expectation
values of general normal ordered operator functions agrees exactly with the
semi-classical prediction combining the quantum theory of matter and the
theory of classical gravitational waves. This {\em in most cases is the
general relativistic prediction itself}, since the quantum nature of matter
is not important in cases of coherent detection. \ The operator for the
intensity of detected waves, for example, has the form $\hat{E}^{(-)}\hat{E}%
^{(+)},$ where $\hat{E}^{(+)}$ is the positive frequency part of the
quantized gravitational radiation field, and the relevant operator function $%
a^{\dagger }a$ is normal ordered. \ The first order and second order
coherence functions, convenient for characterizing the quantum nature of the
radiation, are also normal ordered operator functions and what we already
know about `classical' coherence theory has to be matched exactly by a
theory of quantum gravity. \ This is a powerful and useful indicator to the
theory of quantum gravity.

4) Correspondence principle: The general formulation of quantum
gravitational optics in terms of the diagonal coherent state representation
is exact and does not invoke approximations that depend on the relative
magnitude of physical quantities in comparison with either the Planck
quantities or the quantum mechanical action involved in the problem. Our
demonstration that a quantum coherent state is radiated from a single
gravitationally bound massive binary system allows one to see the approach
towards classical correspondence clearly, maintaining consistency of the
applicability of quantum theory to large and massive systems. \ However, the
matter sector, with its `successful' quantum theory, usually needs such
comparisons to talk about correspondence with the classical world. \ In this
sense, the gravitational radiation sector seems to be fully quantum
mechanical already and the problems that we face today to formulate a
plausible theory of quantum gravity might entirely be in the matter sector!

Having highlighted the main features of quantum gravitational optics, we
suggest that a fresh look at constructing the theory of quantum gravity may
be taken starting from these reliable and fully quantum gravitational
results in the radiation sector of gravity.

\end{document}